\documentstyle[twoside,fleqn,espcrc2,fps]{article}

\newcommand{\AmS}{{\protect\the\textfont2
  A\kern-.1667em\lower.5ex\hbox{M}\kern-.125emS}}

\title{Scaling of gauge balls and static potential in the confinement
  phase of the pure U(1) lattice gauge theory}

\author{J. Cox$^{\rm a}$\thanks{speaker at the conference},
  W. Franzki$^{\rm a}$,
  J.~Jers{\'a}k\address{Institut f{\"u}r
    Theoretische Physik E, RWTH Aachen, Germany}, 
  C. B. Lang\address{Institut f{\"u}r Theoretische Physik,
    Karl-Franzens-Universit\"at Graz, Austria},
  T. Neuhaus\address{ZIF, Universit{\"a}t Bielefeld, Germany},
  A. Seyfried\address{BUGH Wuppertal, Germany} and
  P. W. Stephenson\address{DESY, Zeuthen, Germany}
  }

\begin{document}

\begin{abstract}
We investigate the scaling behaviour of gauge-ball masses
and static potential in the pure U(1) lattice gauge 
theory on toroidal lattices. An extended gauge field action
$-\sum_P(\beta \cos\Theta_P + \gamma \cos2\Theta_P)$ is used with $\gamma= -0.2$ and 
$-0.5$. Gauge-ball correlation functions with all possible 
lattice quantum numbers are calculated. Most gauge-ball masses 
scale with the non-Gaussian exponent $\nu_{\rm ng}\approx 0.36$.
The $A_1^{++}$ 
gauge-ball mass scales with the Gaussian value $\nu_{\rm g} \approx 0.5$
in the investigated range 
of correlation lengths. The static potential is examined 
with Sommer's method. The long range part scales 
consistently with $\nu_{\rm ng}$ but the short 
range part tends to yield smaller values of $\nu$.
The $\beta$-function, having a UV stable zero, is obtained from the running
coupling. These results hold for both $\gamma$ values, 
supporting universality. 
Consequences for the continuum limit 
of the theory are discussed.
\end{abstract}

\maketitle

\section{INTRODUCTION}
Simulations of the pure U(1) lattice gauge theory
on sphere-like lattices \cite{JeLa96,LaPe96} strongly suggest that its phase
transition is of second order for a suitable choice of the action and that the
critical exponent $\nu$ has a non-Gaussian value $\nu_{\rm ng}\approx 0.36$.
The inclusion of a monopole term to the Wilson action on toroidal lattices
also seems to produce a clear second order transition with a (different)
non-Gaussian value of $\nu$ \cite{KeRe97b}.
A non-Gaussian ultraviolet fixed-point in
four dimensional quantum field theory
is an interesting situation, worth
of further examination. Also speculations about a connection
with softly broken $N=2$ supersymmetric Yang-Mills theory \cite{AmEs97}
motivate a more detailed numerical investigation of the 
continuum limit(s) of the lattice model.

To learn more about the corresponding quantum field theory we study
the gauge-ball correlation functions and Wilson loops. As this
would be a tricky task
on a sphere-like lattice, our simulations are performed on the
torus. We avoid the metastability ---
present on small tori even at $\gamma=-0.5$ --- by keeping sufficient distance to
the transition. This restricts the largest correlation length we can achieve to values
below 5.

In \cite{CoFr97b} we calculated gauge-ball masses and the static potential on
toroidal lattices with the extended Wilson action
\begin{equation}
S = -\beta \sum_P \cos\Theta_P  - \gamma \sum_P\cos 2\Theta_P
\end{equation}
at $\gamma=-0.2$. Both the gauge balls and a preliminary investigation of the
long range part of the potential confirm the nontrivial scaling in the
confinement phase with
$\nu_{\rm ng}$, except for the $A_1^{++}$ gauge-ball ($0^{++}$ in
the continuum), which scales with $\nu_{\rm g} \approx 0.5$. Now we repeat the simulation
at $\gamma=-0.5$ to verify universality.
We also make a preliminary analysis of the short-range part of the potential.
\section{GAUGE-BALLS}
\begin{figure}[tbp]
  \begin{center}
    \leavevmode
    \fpsysize=7.0cm
    \fpsbox[70 70 530 770]{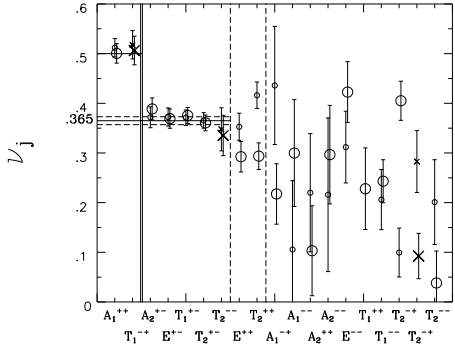}
  \end{center}
  \vspace{-1.5cm}
  \caption{$\nu$ resulting from a fit according to $m_j = c_j
    (\beta_c-\beta)^{\nu_j}$ with a common $\beta_c$. Large (small) symbols correspond
    to $\gamma=-0.5$ ($\gamma=-0.2$). Crosses have only $\vec{p}\neq 0$ and
    can mix with other quantum numbers.}
  \label{fig:nu_gb}
    \vspace{-0.4cm}
\end{figure}

We have investigated the correlation functions in all lattice symmetry sectors
$R^{PC}$ ($R= A_1, A_2, E, T_1, T_2$).
The preliminary results at $\gamma=-0.5$ are very similar to
those at $\gamma=-0.2$ \cite{CoFr97b}:
\begin{itemize}
\item except for $A_1^{++}$, all gauge-balls are consistent with scaling with
  $\nu_{\rm ng}=0.365(8)$ (see Fig. \ref{fig:nu_gb})
\item $A_1^{++}$ scales with $\nu_{\rm g}\approx 0.5$
  and thus becomes presumably massless scalar in the non-Gaussian continuum
  limit
\item the energy spectrum of the non-Gaussian states is compatible with
  being integer multiples of a lowest energy. Thus it may consist of
  n-particle-states of a fundamental object. The one-particle states are 
  $R^{+-}$ with $R=A_2, E, T_1, T_2$ of mass $m_{\rm ng}$.  
\end{itemize}
At $\gamma=-0.2$ the largest lattice size used was $20^340$ which restricted the
(scalar) correlation length to $\xi\leq 3$ in order to avoid phase flips.
At $\gamma=-0.5$ we
use lattices up to $32^364$ allowing for $\xi\leq 5$.

\section{STATIC POTENTIAL}
The static potential is obtained in a standard manner from the exponential
falloff of the expectation value of rectangular Wilson loops of spatial extent $R$ and
temporal extent $T$:
\begin{equation}
  \left\langle W(R,T)  \right\rangle \propto \exp\left( -V(R)T \right)
\end{equation}
Small $T$ values that seemed to be contaminated with higher exitations and larger
$T$'s with a too large relative error were excluded from the fits, that determine the
potential $V(R)$. To analyze the potential we
pursued four strategies:
\subsection*{(a) Coulomb plus linear ansatz for all $R$}
At each $\beta$ the static potential is fitted with the ansatz
\begin{equation}
  V(R)=-\alpha_{\rm ren} V_{\rm L}(R)+ \sigma R + C
  \label{V_of_R}
\end{equation}
where $V_{\rm L}(R)$ is the lattice Coulomb potential or the inverse 3D lattice
Laplacian (see \cite{CoFr97b} for details). The string tension obtained in
this way is fitted as a function of $\beta$,
\begin{equation}
  \sigma(\beta)\propto(\beta_{\rm c}-\beta)^{2\nu}.
  \label{sigma_scale}
\end{equation}
This procedure results in a quite
small value for $\nu\approx 0.30$. There is no apparent change of
$\nu$ when going closer to the phase transition which would be a sign of
non-asymptotic behaviour. This behaviour is insensitive to replacing
the Coulomb potential by a Yukava potential with e.g. a gauge-ball mass as a mass parameter.
\subsection*{(b) Straight line fit for large $R$}
The string tension is determined as the slope of a straight line, fitted to the
three points of the potential with the largest accesible $R$. The $\nu$
obtained by a fit acc. to (\ref{sigma_scale}) is compatible with $\nu_{\rm
  ng}$. From this we conclude that the long range part of the
potential (i.e.~$\sqrt\sigma$) scales with the same non-Gaussian exponent as
$m_{\rm ng}$.
\subsection*{(c) Sommer method}
We define a length scale as proposed in \cite{So94}. First, we choose a
dimensionless number $b$  and then define $R_0$ implicitly by
\begin{equation}
  \left. -F(R)R^2\right|_{R=R_0} = b,
\end{equation}
where $F(R)$ is the force at distance $R$.
For fixed but arbitrary $b$, $R_0$ is expected to scale like an inverse mass:
\begin{equation}
  R_0(b,\beta)\propto (\beta_c-\beta)^{-\nu}.
  \label{R0_scale}
\end{equation}
\begin{figure}[tbp]
  \begin{center}
    \leavevmode
    \fpsysize=7.0cm
    \fpsbox[70 70 530 770]{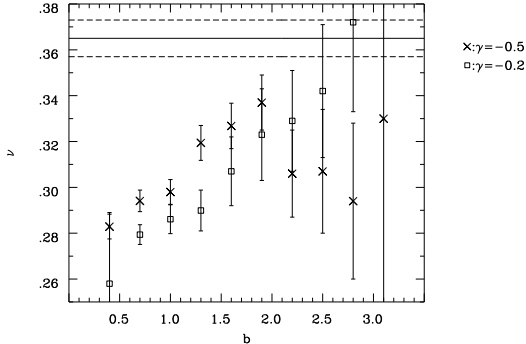}
  \end{center}
  \vspace{-1.5cm}
  \caption{The exponent $\nu$ obtained from a fit of $R_0$ to
    eq. (\ref{R0_scale}). It depends on the 'irrelevant' parameter $b$.}
  \label{fig:nu_b}
    \vspace{-0.4cm}
\end{figure}
The exponent $\nu$ determined by such a fit 
depends on $b$ as shown in Fig. \ref{fig:nu_b}. For small $b$, $\nu$
is considerably smaller than $\nu_{\rm ng}$. It
increases and seems to approach $\nu_{\rm ng}$ for large $b$. This
confirms our earlier findings, that the long range part of the potential
(corresponding to large $b$)
scales with the exponent $\nu_{\rm ng}$,
but that the short range part
differs in that respect.

A possible interpretation is: Define
$R_\epsilon$ to be the distance at which the potential becomes clearly non-linear,
e.g. by calculating the distance where the curvature exceeds a fixed
$\epsilon$. Then the energy scale defined by $R_{\epsilon}^{-1}$ goes to infinity in the
continuum limit with the string tension and the non-Gaussian gauge-ball masses
fixed
to a finite value:
\begin{equation}
  \frac{R_{\epsilon}^{-1}}{m_{\rm ng}}\propto (\beta_c-\beta)^{\nu(b_\epsilon)
      -\nu_{\rm
      ng}}\longrightarrow \infty \quad{\rm if}\quad
  \beta\longrightarrow\beta_c. \nonumber
\end{equation}
The distance where the potential is non-linear thus shrinks in physical
units. However, further study of $R_0$ is required to determine the form of
the potential in the continuum limit.
\subsection*{(d) $\beta$-function}
\begin{figure}[tbp]
  \begin{center}
    \leavevmode
    \fpsysize=7.0cm
    \fpsbox[70 70 530 770]{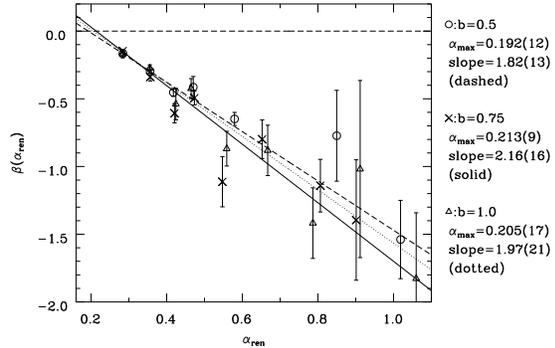}
  \end{center}
  \vspace{-1.5cm}
  \caption{$\beta$-function at three $b$-values: the derivative is taken by a straight line fit to
    five adjacent points in the running coupling plot at each case. Both
    $\gamma$'s were considered. Fixing the slope to 2 results in $\alpha_{\rm
    max}=0.206(5)$.}
  \label{fig:betafunc}
  \vspace{-0.4cm}
\end{figure}
Defining the running coupling like in \cite{So94} (eq. (3.1))
results in a unique
curve --- independent of the parameters $\beta$ and $\gamma$ --- and
calculating the $\beta$-function by taking numerically the
derivative w.r.t.~the log of the scale parameter gives approximatly a straight
line. It has a zero, corresponding to the UV stable non-Gaussian fixed point.
The slope is about 2 in the vicinity of the fixed point --- independent of
$b$. When extrapolated linearly to the Coulomb phase, it roughly agrees with
$\beta$-function values obtained there from the Coulomb potential and the
$A_1^{++}$ resonance energy.
Furthermore, the fixed point value of the running coupling is consistent
with $\alpha_{\rm ren, max} \approx 0.2$ calculated in the Coulomb phase
\cite{JeNe85,CoFr97b}, which is
supposed to be universal.

\end{document}